# COMO*Net*: Community Mobile Network

Primal Wijesekera and Chamath I. Keppitiyagama
University of Colombo School of Computing, Sri Lanka
*primalwijesekera@yahoo.com, chamath@ucsc.cmb.ac.lk*

*Abstract*—The density of mobile phones has increased rapidly in recent years. One drawback of the current mobile telephone technology is that it forces all the calls to go through cellular base stations even if the caller and the callee are within the radio range of each other. Hybrid cellular networks and Unlicensed Mobile Access (UMA) have been proposed as solutions that enable mobile phone users to bypass cellular base stations. However, these technologies either require special hardware or in some cases have to rely on the service providers. We identified that most of the Commodity-off-the-Shelf mobile phones are Wi-Fi (and Bluetooth) enabled. We propose a Community Mobile Network (COMONet) which utilizes Wi-Fi (and Bluetooth) to build ad hoc network among mobile phone users to bypass GSM base stations whenever possible. COMONet does not depend on special non-commodity hardware and it is a software based solution. COMONet monitors all the available paths over the ad hoc network and it transparently switches to the regular path over the service provider's GSM base station if a path is not available over the ad hoc network. In COMONet the caller and the callee do not have to be within the Wi-Fi or Bluetooth range of each other to make a call since the COMONet is capable of routing calls through the other mobile nodes that are participating in the COMONet.

## I. INTRODUCTION

There is an apparent increase in the density of mobile phones and this trend is seen in developed countries as well as developing (third world) countries. In some third world countries, rate of mobile networks expansion has been higher than some developed countries. In the current GSM setup all the calls are forcefully sent through GSM base stations even if the caller and the callee are within the radio range of each other - this situation becomes more and more apparent with the increase of the mobile phone density. Attempts taken so far to overcome this problem situation either require special hardware [10] or in some cases [8] have to rely on the service providers.

Still being a single hop communicator gives us a good example for its old fashion nature. Latest developments in ad Hoc networks, wireless networking and multi hop communication would really change the way we use the mobile phone. One good example would be that a person wants to make a call to a person in the same floor, will not work if either one does not have the coverage of the mobile service provider. That is if either one of them does not have a nearby mobile base station then they cannot communicate even though they are close to each other. Thus forcing to route through a cellular base station would impose restrictions on mobile users.. This becomes more evident in the event of an emergency (such as Tsunami) where all the GSM base stations are destroyed and no one can make phone calls even within a small area.

Resolving above mentioned limitation indirectly eases the dependency of mobile base station. That is the caller would be able to bypass the mobile service provider to make the phone call which is a different mechanism to the way of GSM. There are some notable research projects as well as very few commercial products catering the need to a certain extent, yet they could not achieve how to fully bypass the GSM network.

*A. Solution*

This paper centers on a novel approach which bypasses the mobile base station when making the call. We call this approach as COMONet (Community Mobile Network). The term "Community" symbolizes a major concept in the approach since it utilizes the luxury of high mobile user density to route a voice call to the other party. This feature makes this approach a unique one in its kind. Since it bypasses the mobile base station this leads to a totally new research avenue in mobile computing which is mobile service provider independent mobile communication.

COMONet is a P2P mobile network. As its name suggests what we are trying to achieve is to create a community of mobile users to facilitate voice calls between two parties. As long as both caller and callee are in the community there is no need of a mobile base station. In very broad terms, when a person dials a number COMONet searches whether the dialed mobile phone number is within the reach via ad hoc mode Wi-Fi (or Bluetooth). If the search turns out to be positive the call will be routed via the path in which the intended party was found. Otherwise it will be sent through the mobile service provider. When a call is made through a mobile phone, if the parties are not stationery, it is possible to lose peer connections. As a countermeasure there is a continuous process to monitor the status of the connection currently being used and to check for other alternatives paths. If a better path is found than the current path, the call will be switched to the new path seamlessly without breaking the call. By the time of dialing a number, if there is no path found via COMONet then the call goes through the GSM network. However as the caller moves, if a connection is made with the other party via the community then the call will be automatically switch to the respective path found ending the Mobile Service provider's link. This switching can happen in either direction from Wi-Fi (or Bluetooth) to cellular and from cellular to Wi-Fi (or Bluetooth).

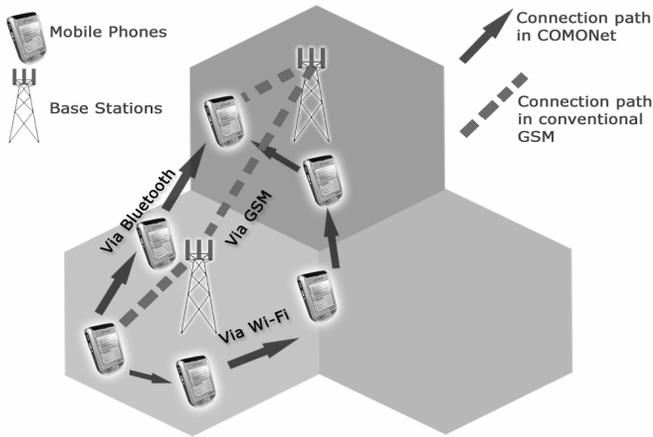

Fig. 1. High Level View of the system

## II. FEASIBILITY

The approach it has taken is an out of the box approach yet there are few critical factors which are beyond the control of COMONet which will decide the feasibility of the project. Feasibility plays such a major role in COMONet since the actual benefits of this approach can be ripped only if it is can be implemented.

### A. Mobile User density

The concept of the COMONet heavily relies on the availability of the people around the two parties who are involved in the conversation.

This is indeed questions the mobile user density where the call is taken. The mobile user density is a luxury left unutilized in the mobile industry. This is a common scenario in all part of the work irrespective of the economy and other factors.

According to a recent research conducted [1], in Germany alone there was a net addition on 5.8 Million new mobile subscribers in 2006 which leads to the penetration level of 103.2%. In USA, net addition was nearly 25 Million new subscribers. Even in countries like Pakistan and Bangladesh annual net addition is over 10 Million new subscribers.

It is clearly shown that there some countries where mobile subscriber penetration is even more than 100% (it is due to the usage multiple service providers). Another important fact that is being presented by this table is that the amount of new subscribers that is added each year. Mentioning about the mobile user density per $Km^2$, in UK it is 268 users, in Hong Kong its 6431 and in New York its 240 users.

The above mentioned facts shows that finding a mobile user in between two end parties is not that hard and the rapid growth of the no mobile subscribers will leave no doubt on the availability of mobile users when a call is made over the COMONet.

### B. Capability of intermediate nodes

If a mobile phone to communicate directly with another mobile phone, then that mobile phone has to be equipped with either WLAN (Wi-Fi) or Bluetooth. Then the question arise will an intermediate mobile phone equip with those features.

In 2005 there were 316 million Bluetooth enabled phones were shipped and it is expected to go high as 866 million (i.e. nearly 90% of entire mobile phone production).

It is just now that Wi-Fi enabled phones are hitting the mass market. In a recent market research conducted, they have found that in countries where WLAN enabled mobile phones are introduced first, there is a tendency that 21% - 25% mobile subscribers are switching to WLAN enabled mobile phones. In a similar research conducted by, they are expecting around 300 million Wi-Fi enabled mobile phones in the near future. What is interesting in these facts are that industry is moving forwards for WLAN enabled mobile phones and more interestingly even the subscribers are willing to switch for a change.

It can be concluded that given a mobile phone is found to route the voice call, the probability of having necessary capabilities in that phone to route the voice call is acceptable and it is likely to increase in the future.

### C. Coverage of COMONet

COMONet is a multi hop solution yet the area that it covers is small compared to conventional GSM cell. The distance comes into play when we are making a call via the COMONet since other party might not be reachable because of the distance. Thus, COMONet will work well for a given community such as a university. Then the question arises how frequent one would make a call to another one in the same premises.

A lecturer in a Nigerian University [3] has done an interesting research on mobile usage patterns of university students. An interesting observation made during the research was that nearly one third (30%) of entire mobile usage was on academic matters.

It shows that there is a high probability that one might call another person who is in the same premises. University is a mere example where COMONet can be used effectively. Other places where COMONet can be used would be an office environment, shopping mall.

### D. User willingness towards the mobile community

The most important factor in deciding the practicality of COMONet is the willingness of the people towards the community. If no one is willing to participate in the community then the whole concept of COMONet is over. A survey was conducted to find on the willingness and the user perception on COMONet.

In the survey it was asked on the extent they would like to corporate with the community and the extent of the impact that COMONet would have on their routine mobile usage. The summarized results were shown below.

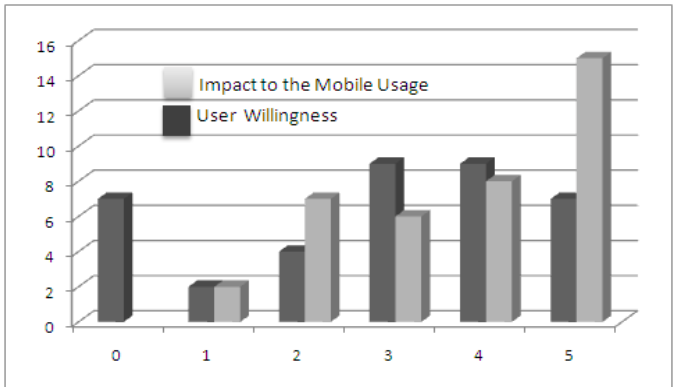

Fig. 2. User Willingness and Usage Impact

The Figure 2 shows that even though all of them think this has a huge impact in reducing their mobile bills yet few are not fully prepared to let their phones to be used as mobile routers. This perception is more likely to change in the future as they will gain more benefits than any drawbacks.

### III. RELATED WORK

Lin et al.[4], propose a new architecture called Multi-hop Cellular Networks (MCN) as opposed to Single-hop Cellular Networks (SCN). MCN is developed combining features in SCN and ad hoc networks. In SCNs relaying is done only through the GSM base station but in MCN it is extended so that even mobile stations are involved in relaying packets. It always connects to a mobile base station through multiple hops.

Ananthapadmanabha et al.[5] proposed an architecture, where all the routing decisions are taken by the base station which leaves a high dependency on the base station. Another architecture is proposed by Luo et al[6]. It is called as UCAN: A Unified Cellular and Ad Hoc Network Architecture. If it finds mobile phones with weaker signals, then it finds an intermediate mobile node through which it routes the signals to the intended party. Similar approaches were proposed by Wu et al.[7] called iCAR. Again there is a high dependency on the mobile base station compared with the COMONet.

Another instance where the cellular network is merged with wireless networks is UMA (Unlicensed Mobile Access) [8] which mostly concentrate on the last mile connection in GSM. If one have a wireless network at home it enables the mobile phone to connect to the wireless network at home and through that to the mobile service provider. In COMONet, if two parties are in near proximate it will try to connect to each other without using the GSM base station and COMONet can be implemented using commodity off the shelf cellular phones.

Moving into another research being done, Luc André Burdet [9] has done a research on "VoIP on a Symbian Mobile Phone". In which he has made a VOIP call via Bluetooth to another peer thus it can be routed even via Internet to the other party. In September, 2007 a Swedish company called TerraNet [10] has published information about its proprietary solution where a direct link can be made between two specifically designed mobile phones. It routes voice through a mesh network thus there is no need of a mobile service provider but all the other details about the architecture is not publicly available. Still both parties need to have the specific mobile phone to work in TerraNet as oppose to COMONet where it can be implemented using commodity off the shelf cellular phones.

What makes COMONet a unique solution in the domain is, it overcomes all of the above mentioned limitations to great extent and more importantly to implement COMONet there is no need of specific hardware requirements. It can be implemented using COTS cellular phones.

### IV. DESIGN

COMONet is a distributed system, running in several mobile nodes. Each node in COMONet can be several states in a given time period. Namely a node can be a in a conversation, might be routing data for another party, might be searching for a path or can be in all those states simultaneously. This is called as multiple views of the architecture in COMONet.

Further in a given view, it can be grouped into three different layers based on their main contribution towards the system as whole. The upper most layers is the Audio Proxy layer, in which the core functionality is to read the voice from the microphone in the phone and write the voice data into the ear piece of the phone.

The intermediate layer is the Data layer. The main functionality of this layer is to handle data streamed via the ad hoc network. At any given time, a particular node in the network can be sending data to the network or receiving data via streaming or routing data from one node to another. All of these tasks are handled by this layer.

The last layer in the system is the connection layer. It deals with all the tasks with the current active connection, searching for a new connection and the monitoring the current active connection. The connection layer can be in three different modes; first one would be the WLAN mode, in which the network is formed via connecting ad hoc mode Wi-Fi. Another option would be connecting each other via a Bluetooth connection and at last it can be connected via a conventional GSM network. Switching between these connections are happening in this layer and the important feature is the last two layers are responsible for making all these switching transparent to the upper most layer which is a vital factor towards the seamless transfer of data between two end points without breaking the voice call.

If the functionality of the COMONet is considered as a whole, it first check for any path via the COMONet leading to the other party, then it has to stream the data via the selected path and it has to continuously monitor the current path.. This can be further divide into two parts based on the medium on which it functions on, namely wireless or conventional GSM. Another view is the monitoring view where it continuously monitors the current connection status and look for alternative paths. Finally the control view which helps to locate the other party via the community.

## A. Wireless View

This is the main view functioning in the architecture this view is activated when the audio is streamed over a wireless connection to the other end. In this there are three components are activated namely Audi Proxy Server (APS), RTP Session, W-Connection. APS access the microphone and read the voice and at the same time write data to earpiece of the phone.

Next main component in this view is the RTP Session, whose main functionality is to read data from APS and generates RTP packets and send it to the other party. At the other end this will read the RTP packers received and write back to the APS which will be later read by the APS.

In COMONet, no SIP or SDP protocols are involved. It was due to the fact that the whole idea behind COMONet is to get away from a centralized environment.

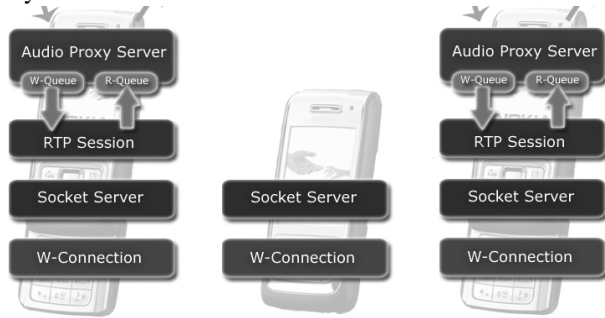

Fig. 3. Wireless View of COMONet

## B. GSM View

This is view is activated when W-Connection layer is not in a position to form a connection with the other intended party. Upper most layers are same as it was in the earlier view as well. But the underlying layer is different to previous one. In GSM view data is transferred to the other party via a 3G connection.

## C. Protocol manager

The routing protocol designed in COMONet, is inspired by the mechanism used in Ad Hoc On-Demand Distance Vector Routing Protocol (AODV). It has few modifications to original routing protocol.

*1) Node discovery*
- When a node wants to communicate with another node it broadcast a path request.
- When another node receives a path request first it checks whether it matches with its own IP,
  o If so then send a path reply to the source address.
  o If the current node is already routing data for the requested IP then, reply to the request.
  o If it is not matched then rebroadcast the path request after decreasing the hop count.

In initiating a path requests, maximum hope count one could go is 7.

*2) Path switching*

In the event of a path loss, the data loss should be minimal. To minimize the data loss which could happen due to a path loss, a mechanism should be placed to detect the fading link before it actually breaks out.
- Initiator of the connection send heart beat messages to its neighbor in the path.
- If a heart beat message is not acknowledged for a defined length of time, then it assumes there is a path loss and sends a path request. It will then notify its predecessors in the path as well.

## D. Eliminating the Dependency on the Mobile Service Provider

If the routing protocol works irrespective of the mobile service provider then the system will achieve its goal. If there is a unique IP for each and every mobile phone then the routing protocol will work without conflicts.

By using subnet masking we can simply convert the mobile phone number (which is already unique) to a unique IP address. COMONet will be only dealing with the assigned IP address and COMONet can communicate with each other irrespective of the underlying mobile service provider.

## V. IMPLEMENTATION

### A. Underlying OS

The underlying OS used for development was Symbian v9.1. Application is built using Symbian C++.

### B. Thread Programming Vs Event Programming

Thread programming was not recommended in Symbian OS. Even though there are several simultaneously running tasks, different event handlers called Active Objects are in place for each event. These active objects work asynchronously.

### C. Major developed components

Complying with the designed presented earlier in the paper, there are clearly defined set of components in COMONet.

*1) Audio Proxy Layer*

The main functionality of this component is manipulating Audio data dealing with the earpiece and the microphone. It is implemented using APS in Symbian. It a wrapper to the S60 sound device.

*2) RTP Session*

The main functionality of this component is stream audio over the COMONet. Once the destination IP address is found RTP session is started and when there is a change in the destination address the RTP session is modified. If a new connection is made different from the current active connection then a new RTP session has to be initialized but it will not obstruct the APS component.

*3) Routing Protocol*

Main objective of this component is to implement the routing protocol mentioned in the design chapter. The core functionalities of the protocol are Send a path request, Listen to a path request, Listen to a path Reply and Send a path reply

Above mentioned core functionalizes are implemented in separate Active Objects, enabling to run simultaneously. They all are connected to one Protocol Manager which has all the data about the requested IPs, respective replied IPs, etc. Asynchronous feature helps a lot to find another route without obstructing the current voice streaming over the WLAN.

*4) Assigning a IP*

The mobile phone number is used to assign a unique IP for a mobile phone. Divide the number into groups having two digits. First group can be deleted since it is common to all numbers. Eg:

```
07  73  03  14  70   →0773031470
       (128+73).03.14.70
0773031470      →201.03.14.70
```

## VI. EVALUATION

### A. Quality of Service in COMONet

The main parameters which affect the quality of the voice stream irrespective of the model and the nature of the network are as Jitter, delay and packet loss ratio.

Data presented in Table 4 is the mean value of the 10 tests conducted aiming at gathering data. Busy intermediate means, it is having a conventional GSM call. The last column is having the recommended values set by ITU-T under Y.1541 [11] recommendation, parameters for Quality of service.

TABLE IV
QOS PARAMETERS

|  | **Direct** | **One hop** | **Busy Interm.** | **Recomd** |
|---|---|---|---|---|
| Delay | 110 ms | 110 ms | 110 ms | 100 ms |
| Jitter | 1.229 ms | 1.698 ms | 1.987 ms | 50 ms |
| Packet loss (%) | 0.000 | 0.000 | 0.000 | .001 |

In analyzing the above table it is obvious that COMONet has a superior quality of service even when the path is multi hop.

### B. Calculating MOS

MOS – CQS is defined in Methods for subjective determination of transmission quality ITU-T Recommendation in P.800 [12].

*1) Testing procedure*

Test environment and the procedure were done according to the instructions given in the above mentioned recommendations [12].

Following observations are seen after analyzing the data gathered from the above using hypothesis test.

- Voice is not degraded by using the multi hop connection.
- Gender of the speaker has not made any impact.
- Gender of the listener has an impact towards the quality.

The average female MOS score is 3.909 and the average male MOS score is 4.4965. The final MOS score is 4.20275.

### C. Call setup time

TABLE V
CALL SET UP TIME

| (In Seconds) | **Direct** | **One hop** | **Busy Inter.** |
|---|---|---|---|
| GSM | 10 | 10 | 10 |
| COMONet | 1.67 | 2.36 | 2.63 |

## VII. FUTURE WORKS

Community mobile network being the first ever approach in its own nature is just a doorway to many avenues.

### A. Routing Video

This can be easily extended to route the live video images of the parties involved in the call as in conventional video call.

### B. Sending SMS

It will just send the SMS to the community and a given party will deliver the SMS to the intended party as soon as it detects the intended party in the community.

### C. Teleconferencing over COMONet

COMONet can be extended to facilitate several people engaging in the same conversation.

### D. Mobile Social Networking

System such as Facebook can be developed on top COMONet, connecting nearby people in the community.

## VIII. Conclusion

In a nutshell, Community Mobile Network (COMONet) means a way of communicating with people without going through the conventional GSM network. In achieving that mechanisms were defined to find the needed mobile users and cope with the mobility of users.

The voice quality is in the range of acceptable quality. Still projects such as COMONet cannot be proven as successful by just set of numbers but lot of sociological factors affects the success of the project because concepts like community cannot be easily proven with just numbers.